%% file: main.tex
\documentclass[letterpaper,twocolumn,10pt]{article}
\usepackage{usenix2019_v3}

\usepackage{tikz}
\usepackage{amsmath}

\usepackage{cite}
\usepackage{amsmath,amssymb,amsfonts}
\usepackage{centernot}
\usepackage{graphicx}
\usepackage{textcomp}
\usepackage{subfigure}
\usepackage{listings}
\usepackage{array}
\usepackage[htt]{hyphenat}
\usepackage{multirow}
\usepackage{adjustbox}
\usepackage{mathtools}
\usepackage{algorithm}
\usepackage[noend]{algpseudocode}
\usepackage{longtable}
\usepackage{framed}
\usepackage[most]{tcolorbox}

\usepackage{url}

\usepackage{breakurl}

\newcolumntype{C}[1]{>{\centering\let\newline\\\arraybackslash\hspace{0pt}}m{#1}}

\definecolor{commentgreen}{rgb}{0.3,0.5,0.5}
\definecolor{keyred}{rgb}{0.63,0.129,0.258}
\definecolor{codegray}{rgb}{0.5,0.5,0.5}
\definecolor{codepurple}{rgb}{0.58,0.4,0.82}
\definecolor{backcolour}{rgb}{0.95,0.95,0.92}
\definecolor{maroon}{cmyk}{0,0.87,0.68,0.32}

\lstdefinestyle{mystyle}{
    commentstyle=\itshape\color{commentgreen},
    keywordstyle=\bfseries\color{keyred},
    numberstyle=\tiny,
    stringstyle=\color{codepurple},
    basicstyle=\ttfamily\footnotesize,
    breakatwhitespace=false,         
    breaklines=true,                 
    captionpos=b,                    
    keepspaces=true,                 
    numbers=left,                    
    numbersep=5pt,                  
    showspaces=false,                
    showstringspaces=false,
    showtabs=false,                  
    tabsize=2,
    frame={bottomline}
}
\def\BibTeX{{\rm B\kern-.05em{\sc i\kern-.025em b}\kern-.08em
    T\kern-.1667em\lower.7ex\hbox{E}\kern-.125emX}}

\newcommand{\tool}{\textsc{EOSafe}}
\newcommand{\engine}{Engine}
\newcommand{\scanner}{Scanner}
\newcommand{\emulator}{Emulator}

\begin{document}

\setlength{\abovedisplayskip}{4pt}
\setlength{\belowdisplayskip}{4pt}

\date{}

\title{Security Analysis of EOSIO Smart Contracts}

\author{
{\rm Ningyu He}\\
Peking University
\and
{\rm Ruiyi Zhang}\\
PeckShield, Inc.
\and
{\rm Lei Wu}\\
Zhejiang University
\and
{\rm Haoyu Wang*}\\
Beijing University of Posts and Telecommunications
\and
{\rm Xiapu Luo}\\
The Hong Kong Polytechnic University
\and
{\rm Yao Guo*}\\
Peking University
\and
{\rm Ting Yu}\\
Qatar Computing Research Institute
\and
{\rm Xuxian Jiang}\\
PeckShield, Inc
} 

\maketitle

\begin{abstract}
The EOSIO blockchain, one of the representative Delegated Proof-of-Stake (DPoS) blockchain platforms, has grown rapidly recently. Meanwhile, a number of vulnerabilities and high-profile attacks against top EOSIO DApps and their smart contracts have also been discovered and observed in the wild, resulting in serious financial damages. Most of the EOSIO smart contracts are not open-sourced and they are typically compiled to WebAssembly (Wasm) bytecode, thus making it challenging to analyze and detect the presence of possible vulnerabilities. In this paper, we propose {\tool}, the first static analysis framework that can be used to automatically detect vulnerabilities in EOSIO smart contracts at the bytecode level. Our framework includes a practical symbolic execution engine for Wasm, a customized library emulator for EOSIO smart contracts, and four heuristics-driven detectors to identify the presence of four most popular vulnerabilities in EOSIO smart contracts. Experimental results suggest that {\tool} achieves promising results in detecting vulnerabilities, with an F1-measure of 98\%. We have applied {\tool} to all active 53,666 smart contracts in the ecosystem (as of November 15, 2019). Our results show that over 25\% of the smart contracts are vulnerable. We further analyze possible exploitation attempts on these vulnerable smart contracts and identify 48 in-the-wild attacks (25 of them have been confirmed by DApp developers), which have resulted in financial loss of at least 1.7 million USD.\end{abstract}

\input{Section-Intro.tex}

\input{Section-Background.tex}

\input{Section-Vulnerability.tex}

\input{Section-Challenge.tex}

\input{Section-Tool.tex}

\input{Section-Implementation-Evaluation.tex}

\input{Section-Discussion}

\input{Section-Related.tex}

\vspace{-0.1in}
\section{Conclusion}

To the best of our knowledge, this paper presents the first work on detecting security vulnerabilities in EOSIO smart contracts. We propose {\tool}, an accurate and scalable framework, which is capable of detecting EOSIO specific vulnerabilities. Experiment results suggest the promising performance of {\tool}. Our large-scale measurement study further reveals serious security issues in the ecosystem, i.e., over 25\% of the smart contracts are vulnerable and a number of high-profile attacks have been successfully carried out.

\bibliographystyle{IEEEtran}
\bibliography{citation.bib}

\vspace{-0.3in}
\input{Section-Appendix.tex}

\end{document}

%% file: Section-Intro.tex
\section{Introduction}
\label{sec:intro}

With the growing prosperity of cryptocurrencies (e.g., Bitcoin), blockchain techniques have become more attractive and been adopted in a number of areas.
Due to the limited throughput (e.g., Transaction Per Second, aka TPS) derived from the inherent principle of the Proof-of-Work consensus, traditional blockchain platforms (e.g., Bitcoin and Ethereum) cannot be used to support high performance applications. Researchers have proposed different consensus protocols, e.g., Proof-of-Stack (PoS)~\cite{pos} and Delegated Proof-of-Stake (DPoS)~\cite{dpos}, to resolve the performance issues.

As one of the most representative DPoS platforms and the first decentralized operating system, EOSIO has become one of the most active global communities. 
EOSIO adopts a multi-threaded mechanism based on its DPoS consensus protocol, which is capable of achieving millions of TPS. The performance advantage of EOSIO makes it popular for Decentralized Applications (DApps) developers. 
EOSIO has successfully surpassed Ethereum in DApp transactions just three months after its launch in June 2018~\cite{eos-surpass-eth} and has further increased its dominance by dozens of times after another several months~\cite{eos-dapp-dominant}. For example, the transaction volume of EOSIO on average is more than a hundred times greater than Ethereum~\cite{eos-dapp-peak}. As of 2019, the total value of on-chain transactions of EOSIO has reached more than 6 billion USD.

A smart contract is a computer protocol that allows users to digitally negotiate an agreement in a convenient and secure way. In contrast to the traditional contract law, the transaction costs of the smart contract are dramatically reduced, and the correctness of its execution is ensured by the consensus protocol. EOSIO smart contracts can be written in C++, which will be compiled to WebAssembly (aka Wasm) and executed in the EOS Virtual Machine (EOS VM). Wasm is a web standard specifying the binary instruction format for a stack-based VM. It can run in modern web browsers and other environments~\cite{wasm-nutshell}.

However, it is not easy to guarantee the security of the implementation of smart contracts, EOSIO in particular. A number of vulnerabilities have been discovered in EOSIO smart contracts, while severe attacks have been observed in the wild, which caused great financial damages. For instance, in fall 2018, a gambling DApp, EOSBet, was attacked twice within just a month~\cite{eosbet-attack-fake-eos, eosbet-attack-fake-receipt} due to \textit{fake EOS} and \textit{fake receipt} vulnerabilities, causing 40,000 and 65,000 EOS losses, respectively.
Therefore, it is necessary to identify the security issues of smart contracts in order to prevent such attacks.

Unfortunately, most smart contracts on EOSIO are not open-sourced, and there are few analysis tools towards analyzing the Wasm bytecode, which makes it more difficult to detect vulnerabilities for EOSIO smart contracts automatically.
Although many efforts have been made to analyze the Ethereum smart contracts~\cite{jiang2018contractfuzzer, tikhomirov2018smartcheck, grech2018madmax, krupp2018teether, torres2018osiris, hildenbrandt2018kevm, luu2016making, he2019characterizing}, none of them, however, can be applied to EOSIO smart contracts, as these two ecosystems are totally different, ranging from the virtual machine, the structure of bytecode, to the types of vulnerabilities.

Specifically, there exists several challenges to analyze EOSIO smart contracts.
First of all, EOS VM is more complicated than Ethereum VM in regard to their instructions, including both quantity and variety. For example, EOS VM supports floating point operations, type conversion and advanced jump instructions like \texttt{br\_table}.
Secondly, compared with Ethereum bytecode, the Wasm bytecode itself is more complicated to analyze due to the multi-level nested structure in functions, which leads to a complicated jump relationship between basic blocks.
Thirdly, most EOSIO vulnerabilities discovered so far are more complicated than previously discovered simple vulnerabilities (e.g., integer overflow). Thus it usually requires more semantic information, e.g., fields of the platform-specific data structure as the indexes, to model and analyze them. For example, to detect the fake EOS vulnerability (described in Section~\ref{sec:vul:fake-eos}), we need to check the specific value of argument \texttt{code} in function \texttt{apply}.

\textbf{This Paper.} 
We have implemented {\tool}, the first systematic static analysis framework for detecting vulnerabilities of EOSIO smart contracts. 
Specifically, we first implement a symbolic execution engine for the Wasm bytecode, and mitigate the inherent path explosion problem by applying a heuristic-guided pruning approach.
Second, to analyze an EOSIO smart contract and simulate its external interactive environment, we implement an emulator to mimic the behaviors of key EOSIO library functions that are crucial in vulnerability detection.
Third, we propose a generic vulnerability detection framework, which allows security analysts to easily implement their own vulnerability detectors as plugins. 
In this work, we have implemented four detectors aiming to detect four high-profile vulnerabilities, including \textit{fake EOS}, \textit{fake receipt}, \textit{rollback} and \textit{missing permission check} (see \S\ref{sec:vul}).

To evaluate the effectiveness of {\tool}, we first manually crafted a benchmark suite including 52 smart contracts, which is composed of vulnerable smart contracts collected from publicly verified attacks and their corresponding patched ones.
Experiment results suggest that {\tool} achieves excellent performance in identifying existing vulnerabilities.
To measure the presence of vulnerabilities in the EOSIO ecosystem, we further applied {\tool} to all the smart contracts in the ecosystem (53,666 in total). Experiment results suggest that security vulnerabilities are prevalent in the EOSIO ecosystem: over 25\% of the smart contracts (including historical versions) are vulnerable, and a large portion of them have not been patched timely.
To further measure the impact of the vulnerabilities, we collect the transaction records (over 2.5 billion transactions in total), and carefully design a set of conservative heuristic strategies to identify attacks targeting these vulnerable smart contracts.
We have identified 48 attacks in total, as well as 183 missing permission check actions. By the time of this writing, 25 attacks have been confirmed by DApp developers, which have already caused the financial loss of over 1.7 million USD.

We make the following main research contributions:

\vspace{-0.1in}
\begin{itemize}
    \item We propose {\tool}, the first systematic static analysis framework for EOSIO smart contracts, which is capable of detecting four kinds of popular vulnerabilities. Experiment results demonstrate that {\tool} achieves excellent performance.
\vspace{-0.1in}
    \item We propose the \textit{valuable-function-centric} detection framework, which is based on our observed vulnerability-specific pruning strategies, to effectively mitigate the path explosion issue.
\vspace{-0.1in}
    \item We apply {\tool} to analyze over 53K EOSIO smart contracts, and perform the first measurement study of the ecosystem. We reveal the severity of the security issues, i.e., over 25\% of the smart contracts have been exposed to the threats introduced by these vulnerabilities.
\vspace{-0.1in}
    \item We have identified 48 attacks and 183 missing permission check actions related to the identified vulnerabilities, which have caused huge financial loss.
    Most of the severe attacks have been confirmed by DApp Teams.
\end{itemize}

%% file: Section-Background.tex
\graphicspath{ {images/} }
\vspace{-0.1in}
\section{Background}
\label{sec:background}

As the first industrial-scale decentralized operating system~\cite{eos}, the EOSIO platform can achieve high performance, i.e., millions of TPS, to efficiently execute complicated DApps. 
The fact that it performs so efficiently is due in large part to the consensus algorithm it uses, i.e., \texttt{DPoS}. Compared to traditional \texttt{PoW} (adopted by Bitcoin and Ethereum), it does not spend vast amount of computing resources on the unnecessary mining process.
We next introduce some key concepts to facilitate the understanding of this work. 

\vspace{-0.1in}
\subsection{Account Management}
\label{sec:background:account}

An \textit{account} in EOSIO is the basic unit to identify an entity. It can trigger transactions to other accounts in EOSIO. Additionally, to ensure account security and prevent identity fraud, EOSIO implements an advanced \textit{permission-based access control system}. Specifically, the account can assign public/private keys to specific actions, and a particular key pair will only be able to execute the corresponding action. By default, an EOSIO account is attached to two public keys: the \textit{owner} key (which specifies the ownership of the account) and the \textit{active} key (which grants access to activities with the account). These two keys authorize two native named permissions: the \textit{owner} and \textit{active} permission, to manage accounts. Apart from the native permissions, EOSIO also allows customized named permissions for advanced account management.

Unlike Ethereum, an \textit{EOSIO smart contract} is not treated as a separate entity. A smart contract is just a snippet of code stored in an account, which makes it easy to explain why a smart contract in EOSIO is \textit{updatable}, rather than something that cannot be changed and destructed freely by the owner.
Therefore, when an account is invoked by another one, the smart contract in which it resides will be responsible for handling the received invocation. In this way, the most critical component in contract is the \textit{dispatcher}, which can dispatch the requests to the corresponding functions. Defined by EOSIO official, the dispatcher in a smart contract is named as \textit{\texttt{apply}} that is shown in Listing~\ref{lst:apply-example}.
The details of implementation and parameter implications are discussed in \S\ref{sec:background:transaction}.

\vspace{-0.1in}
\begin{lstlisting}[language={C++}, caption={An example of dispatcher \texttt{apply}.}, label={lst:apply-example}]
void apply(uint64_t receiver, uint64_t code, uint64_t action) {
  if(action == N(onerror)) {
    check(code == N(eosio), "exception captured");
  }
  auto self = receiver;
  if((code == self || code == N(eosio.token))) {
    switch(action) {
      EOSIO_DISPATCH_HELPER(TYPE, MEMBERS)
    }
  }
}
\end{lstlisting}
\vspace{-0.2in}

\subsection{EOSIO Transactions}
\label{sec:background:transaction}

\begin{figure}[tbp]
\centerline{\includegraphics[width=1\columnwidth]{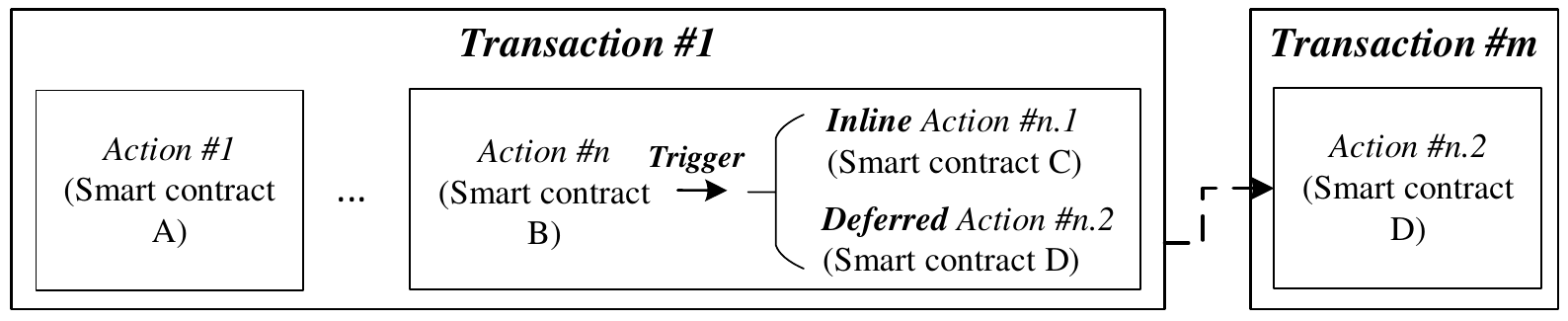}}
\vspace{-0.1in}
\caption{The model of transaction and action in EOSIO.}
\vspace{-0.1in}
\label{fig:tx-action}
\end{figure}

A transaction is the basic unit to be verified by nodes, which is packaged in blocks.
As shown in Fig.~\ref{fig:tx-action}, a \textit{transaction} is composed of one or multiple \textit{action}s. An action is the basic unit to trigger functions. 
For example, the \texttt{action} at line 1 in Listing~\ref{lst:apply-example} specifies the target function name.
An action is responsible for carrying permissions of the invoker. 

An action can trigger other actions under the same context in two ways: \textit{inlined} and \textit{deferred}.
Specifically, the inlined action can be regarded as an ordinary action, which inherits the context of its parent, including the carried permissions. 
As for the deferred action, the reason it was introduced is that the execution time is capped at 30 ms per transaction~\cite{eos-tx-30ms}, and all non-essential actions can be split into deferred actions for execution. Consequently, the deferred action is executed in a different transaction as shown in Fig.~\ref{fig:tx-action}.

\begin{figure}[tbp]
\centerline{\includegraphics[width=\columnwidth]{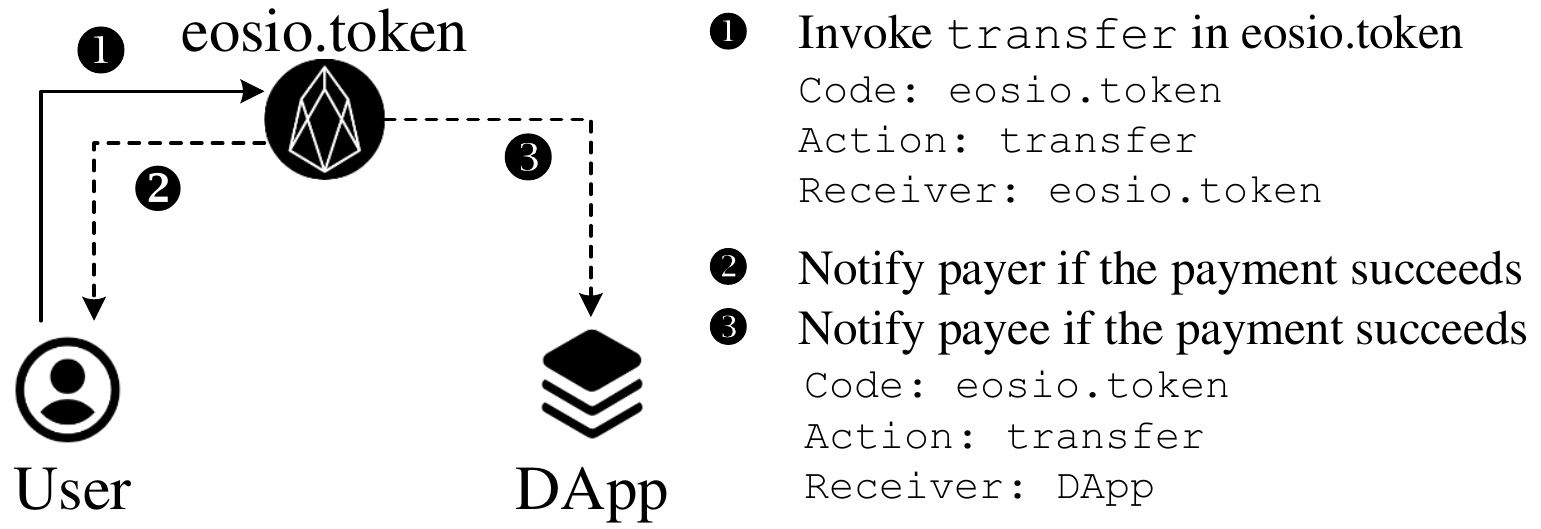}}
\vspace{-0.1in}
\caption{Transferring EOS from the user to a DApp.}
\vspace{-0.2in}
\label{fig:transfer-eos}
\end{figure}

Besides transaction and action, there exists another exclusive mechanism, namely \textit{notification}. 
As shown in Fig.~\ref{fig:transfer-eos}, \textit{EOS} is the official token issued by the account \textit{eosio.token}. It maintains a table to record the holders and their balance. Thus, to transfer EOS to a DApp, a user has to request the \texttt{transfer} function in \textit{eosio.token}. 
For step 1 in Fig.~\ref{fig:transfer-eos}, the \texttt{code}, which indicates whose code is actually invoked, is eosio.token; the \texttt{receiver}, which represents the receiver of the action or notification, is also eosio.token. After updating the balance table, eosio.token will notify both of payer and payee (see step 2 and 3). Note that the \texttt{code} in step 3 is still eosio.token as notification is not an action at all, and the \texttt{receiver} is the DApp. Finally, the notification will also be handled by a dispatcher, just like the action invocation with the same name.
Therefore, we can understand the meaning of parameters used in the dispatcher at line 1 in Listing~\ref{lst:apply-example}.

\vspace{-0.1in}
\subsection{Wasm Bytecode and EOS VM}
\label{sec:background:wasm}
The EOSIO smart contracts are written in C++ and then compiled into WebAssembly (Wasm) bytecode, which will be executed in the EOS VM.
Wasm is a binary instruction format for a stack-based virtual machine. Although it is designed to be an open standard to enable high-performance web applications, it can also be used to support other environments like blockchain.
Due to its efficiency and portability, besides EOSIO, other popular blockchains (e.g., Ethereum 2.0~\cite{ethereum-wasm}) are going to support Wasm.

An EOSIO Wasm binary is called a \textit{module}. Inside a module, numerous \textit{sections} exist. 
To be specific, in \textbf{Function} section, the order of functions is determined, which corresponds to the order of implementation of functions (in low-level instructions) in \textbf{Code} section. All the indexes of functions that appear in \textbf{Element} section can be treated as entries to current module. Additionally, string literals are often used to initialize the \texttt{Memory} section and stored in the \textbf{Data} section.

Like Ethereum VM, EOS VM supports \texttt{Stack}, \texttt{Local}, and \texttt{Global}, which are pushed and popped from a virtual stack by several instructions (such as \texttt{local\_set}, \texttt{global\_get}). Also, EOS VM has an area called \texttt{Memory}, a random-accessible linear array of bytes, which can only be accessed by using specific instructions, e.g., \texttt{load} and \texttt{store}.

%% file: Section-Vulnerability.tex
\section{Vulnerabilities in EOSIO Smart Contracts}
\label{sec:vul}

\begin{figure}[h]
\centerline{\includegraphics[width=0.5\textwidth]{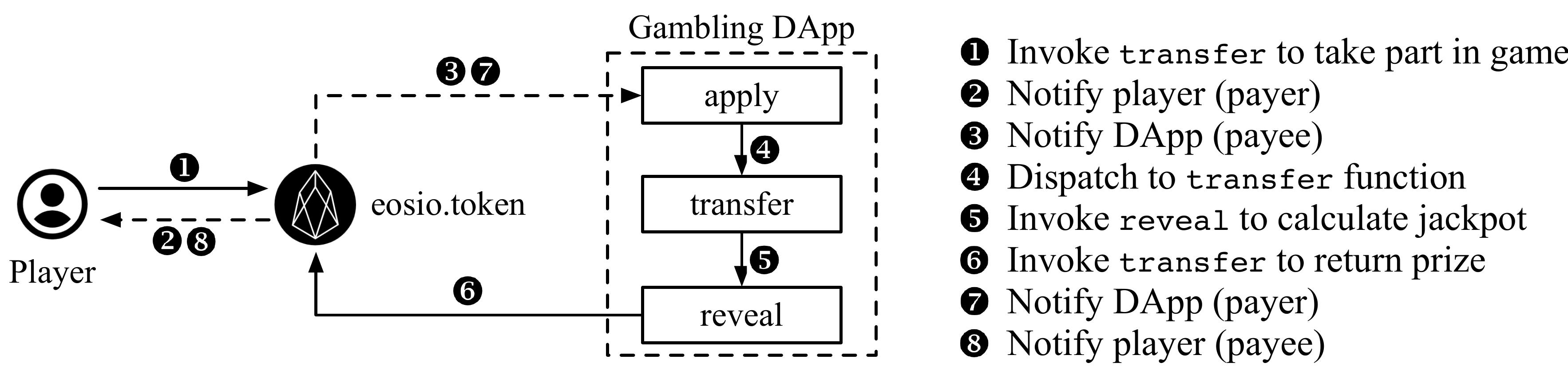}}
\vspace{-0.1in}
\caption{The general life-cycle of smart contract execution.
}
\vspace{-0.2in}
\label{fig:normal-gambling-dapp}
\end{figure}

Attacks can be performed in anytime during the life-cycle of contract execution.
Thus, we first present the general life-cycle of smart contract execution using a gambling DApp as an example, as depicted in Fig.~\ref{fig:normal-gambling-dapp}.
Firstly, the player invokes  \texttt{transfer} in eosio.token to take part in the game. Then, when the DApp receives the notification, it would dispatch the request to \texttt{transfer} through the dispatcher. After that,  \texttt{transfer} would call \texttt{reveal} to calculate a random number to determine if the player hits the jackpot this round. If it is, the DApp will trigger \texttt{transfer} in eosio.token to return the prize to the player.
However, attackers can exploit the vulnerabilities in each step to gain profit.
For example, in steps 3 and 4, failing to rigorously verify the values of the input parameters could be exploited by attackers. On top of that, this whole betting and revealing process has the potential to be maliciously rolled back.
In this section, we discuss four kinds of popular vulnerabilities, relating to the life-cycle of contract execution.

\vspace{-0.1in}
\subsection{Fake EOS}
\label{sec:vul:fake-eos}
Anyone can create and issue a token called \texttt{EOS}, as the token names and symbols are not required to be unique in EOSIO.
Moreover, recall the notification mechanism introduced in \S\ref{sec:background:transaction}, the \texttt{code} in the notification is slightly different from in the action. Therefore, the incorrect verification for \texttt{code} at step 3 in Fig.~\ref{fig:normal-gambling-dapp} may lead to vulnerabilities.

\textbf{Vulnerability Description.}
As the source code of \textit{eosio.token} is entirely public, anyone can make a copy of its source code and issue a token, with the identical name, symbol and code. 
The only difference between the fake EOS and the official one is that they have different issuers.
Therefore, if an attacker transfers the fake EOS to a gambling DApp via the \texttt{transfer} function of the copied contract, the \texttt{code} of the notification received by the project side will not be eosio.token. Additionally, if the DApp happens not to check the value of the \texttt{code}, then the verification in dispatcher will be bypassed.
To mitigate the above issue, some developers narrow down the scope of accepted \texttt{code}, as shown in line 6 of Listing~\ref{lst:apply-example}. ``code == self'' is used to handle the direct call from other accounts, while ``code == N(eosio.token)'' only accepts the notification from the official account. However, due to  short-circuit evaluation~\cite{short-circuit}, if an attacker directly calls \texttt{transfer} in DApp, the verification will also be invalid because there is even no change of balance for both entities. 

As these two cases are only related to fake EOS tokens, in this work, we name both of them as \textit{fake EOS} vulnerabilities.

\subsection{Fake Receipt}
\label{sec:vul:fake-receipt}

If the DApp developer performs a comprehensive check against the \texttt{code}, the notification will then be forwarded by the dispatcher to \texttt{transfer}, as shown in step 4 in Fig.~\ref{fig:normal-gambling-dapp}. However, if the developer does not perform a verification in this step, the DApp can be attacked.

\textbf{Vulnerability Description.}
It is necessary to emphasize 
that the notification can be forwarded, and the \texttt{code} will not change. 
Therefore, DApp might be deceived by the attacker that plays the dual roles (accounts) of an \textit{initiator} and an \textit{accomplice} at the same time.
To be specific, the initiator invokes a regular transfer to accomplice (indicated by \texttt{to}, the argument of the \texttt{transfer} function) through eosio.token. When accomplice is notified by eosio.token, it will immediately forward the notification to DApp without modification.
In this way, the \texttt{code} is not changed, which is the official issuer: eosio.token. Therefore, the dispatcher will be unaware of any anomalies. However, if the parameter \texttt{to} is not checked in \texttt{transfer}, the DApp will be fooled as the token transferring is completed between two accounts that are controlled by the attacker. It results in direct financial loss for DApp developers.

As the notification is triggered by \texttt{require\_recipient}, we name this vulnerability \texttt{fake receipt} .

\vspace{-0.1in}
\subsection{Rollback}
\label{sec:vul:rollback}

In Fig.~\ref{fig:normal-gambling-dapp}, \texttt{transfer} and \texttt{reveal} are the key functions. In \texttt{transfer}, DApp handles the bet that is received along with the player's transfer; in \texttt{reveal}, the developer often uses various \textit{on-chain} state values as seeds (e.g., \texttt{current\_time}, which indicates the timestamp when the action is executed) to generate a pseudo-random number\footnote{The ``pseudo'' is due to all these seeds value are deterministic for lack of a true randomness source on blockchain temporarily.} and finally obtains the result by comparing the generated number with the player's input. 
Note that, in general, the rollback cases can only be found in gambling DApps. We assume the \texttt{reveal} function is always there and is reachable from the entry point (i.e., the \texttt{apply} function) for every gambling DApp.

\textbf{Vulnerability Description.}
Even if the developer does a thorough check on every parameter inputted and checks the caller's permissions before any sensitive actions, a game that matches the model in Fig.~\ref{fig:normal-gambling-dapp} may still be attacked.
To be specific, all the actions are  invoked inline, i.e., locating in a single transaction. Therefore, when the player receives the notification after step 8, he could immediately invoke another inlined action to eosio.token to check his balance. If his balance is reduced, then it means he did not win this round. He can use an assertion statement to force the current action to fail. We have mentioned in \S\ref{sec:background:transaction} that the failure of an action could lead to revert of the whole transaction.
In this way, the player can keep trying until he hits the jackpot. We refer to this malicious rollback as the \textit{rollback} vulnerability.

\subsection{Missing Permission Check}
\label{sec:vul:missing}
Before performing any sensitive operation, the developer should check whether the corresponding permission is carried by the action. For example, before step 5 in Fig.~\ref{fig:normal-gambling-dapp}, the DApp should check whether the caller could represent the actual payer to participate in the game.

\textbf{Vulnerability Description.}
Permission checking is enforced by \texttt{require\_auth(acct)} in EOSIO, which is used to check whether the caller has been authorized by \texttt{acct} to trigger the corresponding function. 
Note that the inlined actions inherit the context of their parents, including the permissions (see \S\ref{sec:background:account}).
Therefore, if an attacker carrying insufficient permission invokes a function, in which it performs sensitive operations via inlined actions and without permission checking, the unexpected behaviors would happen.
We regard all the functions lacking of permission checking as the ones with the \textit{missing permission check} vulnerability.

%% file: Section-Challenge.tex
\section{Technical Challenges and Our Solutions}
\label{sec:challenges}

Take all the factors into account, out goal is to design and implement a symbolic execution based static analysis system to detect vulnerabilities for EOSIO smart contracts. 
To recover more semantic information, we use heuristic-based symbolic execution to perform in-depth analysis. Namely, semantic information will be presented in the constraints generated by symbolic execution along the paths being analyzed. As a result, we are able to use those constraints as patterns to identify vulnerabilities in smart contracts.

\textbf{Comparing with Ethereum Smart Contract Analysis.}
Although there exist a number of static analysis tools proposed for Ethereum smart contracts, it is worth noting that they cannot be applied directly (or even after minor changes) to EOSIO smart contracts due to the differences between the two platforms, including VM models (e.g., allowing global variables), instructions (e.g., supporting floating-point operations) and system-level data structures (e.g., using \textit{multi-index table} to store persistent data).
In brief, these functionalities provided by EOSIO inevitably affect the design/implementation of the proposed system. For instance, we have to consider the side effect caused by the use of external/system libraries (see \S\ref{subsec:challenge-dependency} for details). 
Apart from these differences, the vulnerabilities of EOSIO smart contracts are totally different from those of Ethereum smart contracts, which acquire different kinds of context information to support the detection. 
For example, the rollback vulnerability requires multiple actions being included in one transaction. As such, the detection relies on the propagation of some specific chain state variables (discussed in \S\ref{subsubsec:rollback}).

As a result, \textit{no available symbolic analysis framework could be used to handle the EOSIO Wasm bytecode.}
Specifically, we have to overcome several technical challenges to realize the proposed system.  
On one hand, it is known that symbolic execution based solutions may suffer from inherent shortcomings, \textit{path explosion} in particular. On the other hand, when applied to  vulnerability detection for EOSIO smart contracts, there do exist platform-specific issues, including \textit{memory overlap} and external/system \textit{library dependency}, which will inevitably affect the effectiveness of  symbolic execution. 

\vspace{-0.1in}
\subsection{Path Explosion}
\label{subsec:challenge-path}

In EOSIO, this issue is mainly due to two circumstances: \textit{executing conditional jump instructions} (such as \texttt{br\_if}) or \textit{invoking function call}. Specifically, unlike a normal conditional jump instruction that only generates two new branches, \texttt{br\_table} in EOSIO, however, takes an array whose elements are pointers of destination as the argument. As a result, a single \texttt{br\_table} can lead to $n$ new branches, where $n$ is the length of the array. Apart from those conditional jump instructions, a function call also imposes many new branches to represent all possible callees. Obviously, the number of branches will increase exponentially if there exists a deep call stack. Unfortunately, a concatenation of several deep call stacks is common in EOSIO contracts. As such, there is a practical need to mitigate this issue, otherwise the symbolic execution solution will not be applicable.

To this end, we adopt a \textbf{heuristic-guided pruning approach} to solve the challenge. 
On one side, we rely on several \textit{general} pruning strategies based on our hands-on experience to mitigate the issue derived from branches and deep function calls.
For example, our operational observation suggests that discarding paths under a specific depth threshold, which is determined by the scenario, will not influence the precision of results for (almost) all cases.
Specifically, we expose 
1) an option named \texttt{call depth}, which limits the depth of call stack; 
and 2) an option named \texttt{timeout} for users to limit the process of symbolic execution.

However, the effectiveness of the general mitigation strategies are limited in practice.
Fortunately, 
this issue in EOSIO can be further (partially) resolved when performing vulnerability detection, as we only have to pay attention to some specific features/structures of the vulnerable code snippet. 
For example, when detecting fake EOS and fake receipt vulnerabilities, only \texttt{apply} and \texttt{transfer} functions are taken into consideration. 
All these technical details and \textit{vulnerability-specific} pruning strategies will be discussed in \S\ref{subsec:scanner}.

\begin{figure}[tbp]
\centerline{\includegraphics[width=\columnwidth]{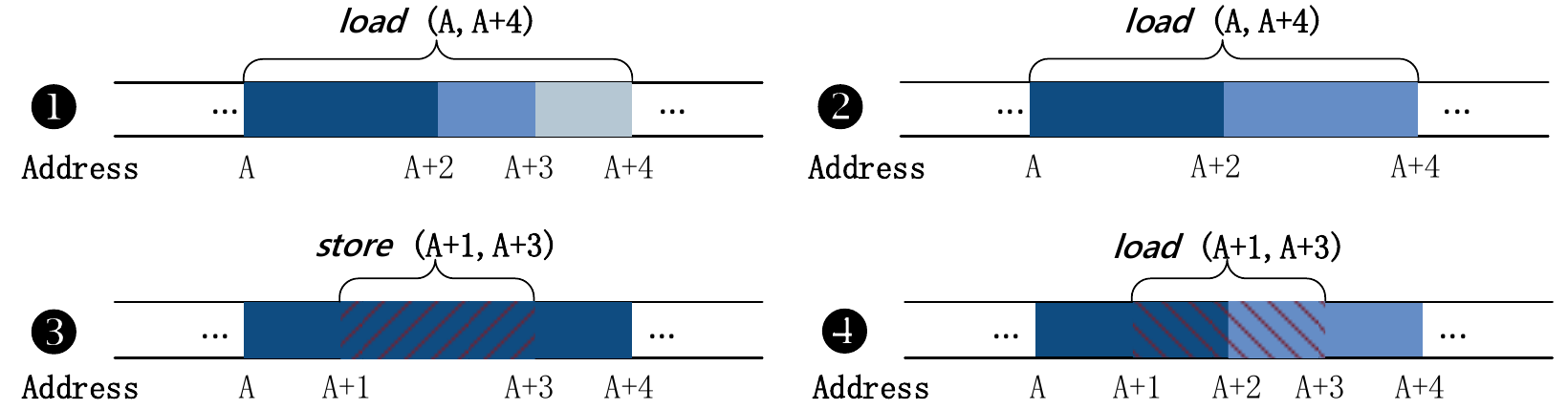}}
\vspace{-0.1in}
\caption{The memory overlap problem.}
\vspace{-0.2in}
\label{fig:memory-challenge}
\end{figure}

\subsection{Memory Overlap}
\label{subsec:challenge-memory}
The memory area of Wasm can be regarded as a vector of uninterpreted bytes~\cite{wasm-memory}, which means users can interpret these raw bits
through \texttt{load} and \texttt{store} with different value types.
The traditional way to emulate memory is to use \textit{linear array}, but it is memory-consuming, due to mimicking the sparse memory layout of the EOSIO smart contract. 
Therefore, we decide to use \textit{key-value} mappings to emulate the memory, where the key is a tuple to specify the address range, and the value is the data being stored, as follows:
$$(\text{lower-bound}, \text{upper-bound})\mapsto \text{data}$$

However, this strategy may lead to the memory overlap (see Fig.~\ref{fig:memory-challenge}). If we use the mapping without optimization, keys $(A+2,A+3)$, $(A+3,A+4)$, and $(A+2,A+4)$ can all be stored directly with no conflicts. As a result, if we retrieve data with the key $(A,A+4)$, there exists two cases meeting the condition (Case 1 and 2 shown in Fig.~\ref{fig:memory-challenge}), which may result in retrieving the wrong data.
Additionally, in Case 3, if we want to update data in $(A+1,A+3)$, we have to traverse the key space to determine if there is only one entity containing the address range we provided, which must be guaranteed to ensure data consistency. In Case 4, we have to concatenate adjacent stored data chunks to determine how to load data in the memory area. In nutshell, all these problems are due to the overlapping memory and improper mapping strategy.

We propose a \textbf{memory-merging} method (see \S\ref{subsubsec:memory_overlap}) to solve the problem by merging allocated memory. 
As aforementioned, Wasm provides over 20 memory access related instructions. 
We will first create key-value mappings for all of the \texttt{store}-related instructions we encountered, where the values are the stored data in bits. 
After that, we are able to handle the cases when the ranges of two keys are adjacent or overlapped according to the proposed \textit{memory-merging algorithm}, which will update the corresponding data chunks to guarantee the precision of execution. 
In brief, we make every effort to guarantee that the interval between any two arbitrary keys is at least one bit.
By doing so, we can successfully overcome the challenge raised in Fig.~\ref{fig:memory-challenge}.

\subsection{Library Dependency}
\label{subsec:challenge-dependency}
To facilitate the development of smart contracts, EOSIO allows the import of external functions as libraries, which means the bodies of these imported functions will not be compiled into Wasm bytecode. EOSIO officially provides plenty of such functions as the system library for DApp developers. They have been widely used in many (if not most) smart contracts. 
As a result, our analysis will be improperly terminated due to the lack of bodies of those imported function calls.  

To resolve the dependency, we have proposed an \textbf{on-demand and semantic-aware} approach (see \S\ref{subsec:emulator}) to emulate the imported functions. 
We only focus on functions whose functionality and side effect are related to our analysis. We have to emulate such functions properly to guarantee the correctness of the final result.
The strength and coverage of the emulation depend on our need to perform the analysis. 
For some functions, we have to cover the arguments, return value and side effect. For instance, the memory-related function, \texttt{memmov}, in which we have to consider all its side effect on the symbolic memory.
For some others, we may only need to consider the possible side effect. For example, for those table-related functions which has no return value and no effect on vulnerability detection, e.g., \texttt{db\_store\_i64}, we can just balance the stack without mimicking its behaviors.

%% file: Section-Tool.tex
\vspace{-0.1in}
\section{System Design}
\label{sec:tool}

\begin{figure}[tbp]
\centerline{\includegraphics[width=\columnwidth]{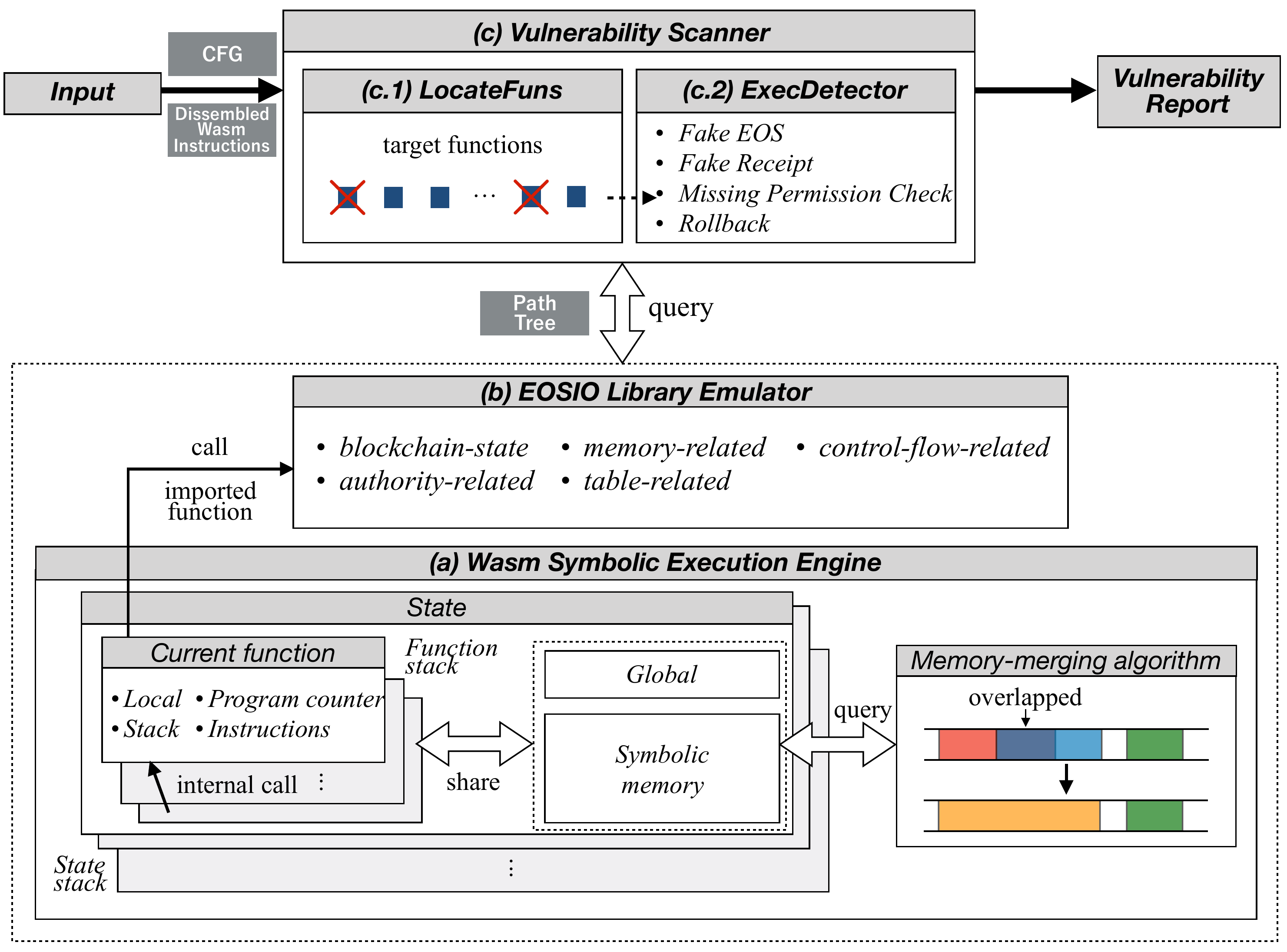}}
\vspace{-0.1in}
\caption{The architecture of \tool.}
\vspace{-0.2in}
\label{fig:architecture}
\end{figure}

Fig.~\ref{fig:architecture} depicts the overall architecture of \tool, which takes the Wasm bytecode of an EOSIO smart contract as the input and eventually determines whether the bytecode is vulnerable.
Specifically, {\tool} is based on \textit{Octopus}~\cite{octopus}, a security analysis framework for Wasm modules, to launch the preprocessing. Each smart contract will be sent to \textit{Octopus} for building its corresponding Control Flow Graph (CFG) with the disassembled Wasm instructions.

{\tool} is mainly composed of three modules, i.e., \textit{Wasm Symbolic Execution Engine} (\textit{\engine} for short), \textit{EOSIO Library Emulator} (\textit{\emulator} for short), and \textit{Vulnerability Scanner} (\textit{\scanner} for short).
As shown in Fig.~\ref{fig:architecture}, the input after preprocessing (CFGs) is fed to the {\scanner} to perform vulnerability detection in a two-step process (\textit{locating suspicious functions} and \textit{detecting vulnerabilities}) with the {\engine} and {\emulator}. Specifically, the {\engine} performs symbolic execution accordingly along with path constraints, which will be used by the {\scanner} to perform vulnerability detection. Additionally, the {\engine} requests {\emulator} to implement the modeled behaviors when the {\engine} encounters the call for imported functions.
Notice that the challenges discussed in \S\ref{subsec:challenge-path} and \S\ref{subsec:challenge-memory} are addressed in \S\ref{subsec:engine} and \S\ref{subsec:scanner}, while the challenge discussed in \S\ref{subsec:challenge-dependency} is addressed in \S\ref{subsec:emulator}.

\subsection{Wasm Symbolic Execution Engine} 
\label{subsec:engine}
The engine is designed as a generic framework to simulate the execution of a smart contract on the stack-based EOS VM. 
It accepts the CFGs and the disassembled Wasm instructions as the input, and symbolically executes instructions within basic blocks in order for all feasible paths. 
During the process, the path constraints are generated accordingly.
Specifically, the module has to maintain two crucial components: \textit{path tree} and \textit{state}.
For the \textit{path tree}, we not only record the constraints generated by symbolic execution, but also all the arguments and return value of imported functions along the path, which contribute to the analysis of vulnerability detection. As to the \textit{state}, we maintain some necessary state-related information, including local/global variables, linear memory, stack, and the subsequent instructions with its corresponding program counter. Specifically, we have addressed the technical challenges mentioned in \S\ref{subsec:challenge-path} and \S\ref{subsec:challenge-memory} as follows.

\subsubsection{Alleviating path explosion with general strategies}
\label{subsubsec:generalstrategies}
We provide two options, including \texttt{call depth} and \texttt{timeout}, for users to mitigate this issue by sacrificing the accuracy.
On one hand, as the name suggests, the option \texttt{call depth} is used to confine the depth of the call stack to prevent the analysis from getting into trouble to deal with complicated branches or deep function calls.
As we know, a single function could have several sets of constraints corresponding to feasible paths within the function, which may lead to an exponential growth of the number of paths. Thus we limit the depth of call stack to improve the coverage.
On the other hand, we may still be in trouble when encountering some cases that are extremely time-consuming.
To guarantee the progress for the whole system, the {\engine} offers another option named \texttt{timeout} to control the maximum execution time for the path-level analysis. Of course, the timeout results will be recorded for further investigation.
Note that, the path explosion issue will be further addressed in the vulnerability scanner (see \S~\ref{subsec:scanner}), as we only have to pay attention to some specific features of the vulnerable code snippets.

\begin{algorithm}[t]
\caption{Memory-merging algorithm.}
\label{algo:memory-merge}
\textbf{Input:} \textit{sm} - symbolic memory, \textit{dest} - insert position, \textit{len} - data length, \textit{data} - data\\
\textbf{Output:} \textit{sm} - updated symbolic memory\\
\textbf{Description:} \textit{es} and \textit{ee} respectively stand for lower-bound and upper-bound of address range of the picked key. The \textit{o} in \textit{os} and \textit{oe} stands for the overlapped.
\begin{algorithmic}[1]

\Procedure{MemoryMerge}{$sm, dest, len, data$}
	\State $isOverlapped, es, ee \leftarrow$ IsOverlapped($sm, dest, len$)
	\If{$\neg isOverlapped$}
		\State $data \leftarrow$ ToLittleEndian($data, len$)
		\State $sm[(dest, dest+len)] \leftarrow data$
	\Else
		\State $os, oe \leftarrow$ CalcOverlap($es, ee, dest, len$)
		\State UpdateOverlappedPart($sm, os, oe, es, ee$)
		\State CatOtherParts($sm, os, oe, es, ee$)
	\EndIf
	\State $keys \leftarrow$ SortKeys($sm$)
	\While{$i+1 < len(keys)$}
		\State $currentKey \leftarrow keys[i]$
		\State $nextKey \leftarrow keys[i+1]$
		\If{$currentKey[1] == nextKey[0]$}
			\State Merge($sm, currentKey, nextKey$)
			\State RemoveKeys($sm, currentKey, nextKey$)
			\State InsertNewKey($sm, currentKey, nextKey$)
		\Else
			\State $i \leftarrow i+1$
		\EndIf
	\EndWhile
	\State \Return $sm$
\EndProcedure
\end{algorithmic}
\end{algorithm}

\subsubsection{Eliminating the memory overlap}
\label{subsubsec:memory_overlap}
We implement a \textit{symbolic memory} to represent the memory of Wasm, and also propose a \textit{memory-merging algorithm} (see Algorithm~\ref{algo:memory-merge}) to emulate the \texttt{store} instruction. 
This algorithm takes the symbolic memory, address, length of data in byte and data as the input, and finally returns a merged symbolic memory without overlapped/adjacent keys as the output. 
Specifically, given a new key that will be stored, we will check whether the address range of an existing key is overlapped with that of the new key or not. If so, the insertion will be performed directly; 
otherwise it will update the overlapped part accordingly and concatenate the non-overlapped parts. 
Furthermore, the keys are sorted in ascending order of the starting position. If two adjacent keys are not overlapped, they will be merged together to form a new key-value pair. 
For example, the existing key-value pairs are:
$$\text{symbolic memory} := \{(0, 2) \mapsto a_0|a_1 , (3, 4)\mapsto a_3\}$$
When $(2, 4) \mapsto a_2|a_3'$ arrives, it will update the overlapped part and concatenate the non-overlapped part on necessary:
$$\text{symbolic memory} := \{(0, 2) \mapsto a_0|a_1 , (2, 4)\mapsto a_2|a_3'\}$$
After that, it will merge the adjacent keys together:
$$\text{symbolic memory} := \{(0, 4) \mapsto a_0|a_1|a_2|a_3'\}$$

In brief, this algorithm guarantees the data consistency by forcing all valid addresses appear only once in the key space. Thus, we can solve all the issues raised in Fig.~\ref{fig:memory-challenge} effectively.

\subsection{EOSIO Library Emulator}
\label{subsec:emulator}
We use the \textit{on-demand and semantic-aware} approach to resolve EOSIO library dependency. 
We have manually analyzed the smart contracts of the top 100 popular DApps and existing known vulnerable smart contracts (see \S\ref{sec:evaluation:rq1:benchmark}) to extract all the imported functions from their \textbf{Function} section (see \S\ref{sec:background:wasm}). 
Then, we classify all the imported functions into five categories according to their main functionalities (as shown in Table~\ref{table:external-functions}) to conduct the emulation.
Lastly, we can retrieve the side effects from the emulated imported functions.
	
\begin{table}[!t]
\caption{The categories of emulated imported functions.}
\centering
\begin{tabular}{cc}
\hline
Category                              & Imported Function Examples  \\ \hline\hline
\multirow{2}{*}{blockchain-state}     & tapos\_block\_num   \\ \cline{2-2} 
                                      & current\_time       \\ \hline
\multirow{2}{*}{memory-related}       & memcpy              \\ \cline{2-2} 
                                      & memmov              \\ \hline
\multirow{2}{*}{control-flow-related} & eosio\_exit         \\ \cline{2-2} 
                                      & eosio\_assert       \\ \hline
\multirow{2}{*}{authority-related}    & require\_auth       \\ \cline{2-2} 
                                      & require\_auth\_2    \\ \hline
\multirow{2}{*}{table-related}        & db\_get\_i64        \\ \cline{2-2} 
                                      & db\_update\_i64     \\ \hline
\end{tabular}
\label{table:external-functions}
\vspace{-0.15in}
\end{table}

Specifically, the emulated imported functions are classified into categories, as shown in Table~\ref{table:external-functions}. The corresponding side effects are summarized in the following.

\textbf{\textit{Blockchain-state functions.}} These functions return constants related to the blockchain system, which are mostly used by the smart contracts as the seeds, to generate the pseudo-random numbers. As they do not introduce any side effect, we just emulate them by directly returning a symbolic value to represent the blockchain state.

\textbf{\textit{Memory-related functions.}} As the name suggests, functions in this category are related to the symbolic memory we have implemented. Therefore, we imitate the behaviors as their original intention, and apply the memory-merging algorithm when inserting the new data. Note that, we throw an exception for undefined behaviors, e.g., the negative length of the \texttt{memcpy} function due to the constraint solving.

\textbf{\textit{Control flow related functions.}} 
These functions are those which may alter or terminate the control flow of a smart contract according to their return results. Therefore, we will fork two paths if necessary. For example, two paths will be generated if the predicate of the \texttt{eosio\_assert} function is a symbolic value rather than a specific boolean value.

\textbf{\textit{Authority-related functions.}} 
As the authority system is merely related to the detection of \textit{missing permission check} vulnerability, we only have to examine the existence of these functions without concerning about the specific permission. Hence, we just return a symbolic value to balance the stack.

\textbf{\textit{Table-related functions.}} 
There is a special data structure in EOSIO that allows for persistent storage of data. Similar to the concept of \textit{storage} in Ethereum, this kind of data is saved on the blockchain that is called \textit{table}. \textit{Table} can be regarded as a database that supports \textit{CRUD} operations (i.e., Create, Retrieve, Update and Delete) by several platform-specific instructions.
For these functions, we only have to focus on the side effects to the memory rather than the internal operations. Specifically, we have implemented them with return values used to update the memory, as follows: 

\vspace{-0.2in}
\begin{gather*}
\text{A} = db\_get\_i64(\text{itr}, \text{data}, \text{length}) \\
i64.store(\text{base}, \text{A})
\end{gather*}

Note that for functions (e.g., \texttt{db\_update\_i64}) that do not have any return value but modify the contents of the \textit{table}, we record their function names and arguments in the constraints.

\subsection{Vulnerability Scanner} 
\label{subsec:scanner}

To detect multiple vulnerabilities, the Scanner is designed as a generic framework to perform the detection. 
It mainly consists of two steps, i.e., \textit{locating suspicious functions} and \textit{detecting vulnerabilities}.
Accordingly, our goal is to realize detectors for the four vulnerabilities introduced in \S\ref{sec:vul}. 

The general strategies proposed in \S\ref{subsubsec:generalstrategies} can alleviate the path explosion problem to some extent, however, it is still not enough to meet our needs.
Fortunately, one key insight can help further mitigate this issue, i.e., we only have to focus on \textit{valuable} functions that call external functions with the ability to change the on-chain state, including \texttt{send\_inline} (see \S\ref{sec:background:transaction}), \texttt{db\_update\_i64} and \texttt{db\_store\_i64} (see \S\ref{subsec:emulator}).
According to our observation, these valuable functions can be heuristically regarded as target functions in most cases, which can significantly reduce the analyzing time. 
As a result, in favor of CFG and path tree (composed of constraints and valuable functions), we can identify vulnerabilities efficiently and accurately.
Specifically, we define the following formulas:

\vspace{-0.2in}
\begin{align}
  \{\texttt{send\_inline}(func)\} &\subseteq con_{X} \label{eq:valuable:1}\\
  \{\texttt{db\_update\_i64}(args, ...)\} &\subseteq con_{X} \label{eq:valuable:2}\\
  \{\texttt{db\_store\_i64}(args, ...)\} &\subseteq con_{X} \label{eq:valuable:3}
\end{align}
Here $con_X$ represents the constraints of a given function, where $X$ is the name of the function being analyzed.
$X$ is said to be a \textit{valuable} function if and only if $con_X$ satisfies at least one of the three criterion.

As a result, the two steps of the detection framework can be further transferred and simplified as a \textit{valuable-function-centric} process: 1) \textit{locating valuable functions}; and 2) \textit{verifying their reachability to launch attacks}. 
Note that the second step of the process is optional since the reachability can always be guaranteed in some cases.
Based on this framework,  we will introduce the details for the four detectors.

\vspace{-0.1in}
\subsubsection{Fake EOS Detection}
\label{subsubsec:fakeeos}
As discussed in \S\ref{sec:vul:fake-eos} and depicted in Fig.~\ref{fig:normal-gambling-dapp}, the fake EOS vulnerability can only be triggered by invoking the \texttt{transfer} function, which becomes the valuable function that can lead to financial losses, which satisfies the following condition: 

\vspace{-0.2in}
\begin{align*}
  \{\texttt{send\_inline}(\dots)\} &\subseteq con_{transfer}
\end{align*}

Moreover, the \texttt{transfer} function must be reachable from the entry (i.e., the \texttt{apply} function) by attackers, which means there does not exist proper verification of \texttt{code} in the \texttt{apply} function. 
Specifically, the detector traverses all the feasible paths generated by symbolically executing the \texttt{apply} function to examine if the constraints of the current path correspond with the following condition:

\vspace{-0.2in}
\begin{align*}
\{action \gets \texttt{transfer}\} \ \cap \ (\{code \gets \texttt{self}\} \ &\cup \\
\{\forall acct \in accounts, code \centernot\gets acct\}) &\subseteq con_{apply}
\end{align*} 

Specifically, it restricts that only the paths associated with the \texttt{transfer} function can be analyzed. 
To accelerate the analysis, the {\engine} will terminate irrelevant paths (if the destination is not \texttt{transfer}) in advance to avoid further execution. 
Then, the detector will examine the value in \texttt{code}, as discussed in \S\ref{sec:vul:fake-eos}. 
Thus, the satisfaction of any of the conditions associated with the \texttt{code} implies the existence of improper verification.
In summary, a smart contract that meets the above conditions is considered to be vulnerable.
\vspace{-0.1in}
\subsubsection{Fake Receipt Detection}
This vulnerability is due to inadequate verification inside the \texttt{transfer} function.
However, the corresponding \texttt{send\_inline} function is extremely deep, which always leads to call depth overflow to break the {\engine}. 
Therefore, it is not feasible to locate the valuable function (i.e., the \texttt{transfer} function) from the entry (i.e., the \texttt{apply} function) through symbolic execution directly.

To solve the problem, we adopt a heuristic-based method instead. 
Specifically, the detector first identifies the \texttt{apply} function, and then enumerates all the relevant basic blocks to verify their jump targets whose indices may point to the suspicious \texttt{transfer} functions. 

After locating the suspicious \texttt{transfer} functions, the detector relies on criteria from \ref{eq:valuable:1} to \ref{eq:valuable:3} to determine the valuable \texttt{transfer} function.
Note that for a given smart contract, there shall exist exactly one \texttt{transfer} function (like Fig.~\ref{fig:normal-gambling-dapp}), which implies that the \texttt{transfer} function is either one of the suspicious functions, or inlined in the \texttt{apply} function, as follows:

\vspace{-0.2in}
\begin{align*}
\exists sus \in set_{suspicious}, \{action \gets \texttt{transfer}\} \subseteq con_{sus}
\end{align*}

or:

\vspace{-0.2in}
\begin{align*}
\forall sus \in set_{suspicious}, \{action \gets \texttt{transfer}\} \centernot\subseteq con_{sus}
\end{align*}

For either of the above two cases, we would further examine the existence of the following protection:

\vspace{-0.2in}
\begin{align*}
func \in \{sus, apply\}, \{to \gets self\} \subseteq con_{func}
\end{align*}
A smart contract that meets the above conditions is regarded to be vulnerable to the fake receipt vulnerability.

Besides, this detector also applies early termination to accelerate the whole process. 
Specifically, for the valuable \texttt{transfer}, the protection should be verified before updating changes for related on-chain states. Thus, it is reasonable to terminate the analysis when encountering any of the three criteria without protection along the path.

\vspace{-0.1in}
\subsubsection{Rollback Detection}
\label{subsubsec:rollback}
As shown in Fig.~\ref{fig:normal-gambling-dapp}, the \texttt{reveal} function often generates random number to determine the jackpot winner, and invokes the \texttt{transfer} function in eosio.token by an inlined action to return the prize. 
Thus, the \texttt{reveal} function becomes the valuable function according to Criterion~\ref{eq:valuable:1}.
In some circumstances, however, computational burden has to be considered when handling the \texttt{reveal} function, i.e., the call depth of the \texttt{send\_inline} function is too deep for the {\engine} to reach.

Fortunately, as it is not necessary to consider the reachability of the \texttt{send\_inline} function in a path for any target gambling DApp (see \S\ref{sec:vul:rollback}), we are able to apply two strategies to accelerate the process to locate the \texttt{reveal} function.
Specifically, the first strategy is to traverse feasible paths on demand. Instead of enumerating all paths, we only examine paths that can be used to resolve the data/variable dependency of the target \texttt{send\_inline} function. 
On the other hand, the second strategy reduces the size of the \textit{path set} being examined by the {\engine} after extracting valuable functions, namely, removing redundant paths whose basic blocks are thoroughly the subset of other paths. 
Consequently, we can achieve the smallest path set to cover  as many basic blocks as possible.

Finally, the detection logic is associated with two properties.
Firstly, our investigation suggests that the \texttt{reveal} function will generate random numbers with the \texttt{rem} instruction along the path inside the constructed path set.
Secondly, as explained in \S\ref{sec:vul:rollback}, if the operands of the modulo calculation are (partially) generated by blockchain-state functions (see \S\ref{subsec:emulator}), the smart contract will be affected by the rollback vulnerability.
In summary, the detection logic must satisfy:

\vspace{-0.2in}
\begin{align*}
\{\texttt{rem}(operand_1, operand_2)\} \subseteq con_{reveal} \  \cap &\\
operand_1 \in \ BlockChainStateFuncs &
\end{align*}
Here $operand_2$ is always a constant or a variable that has nothing to do with the blockchain state.
If the above conditions are met, we can confirm this contract is vulnerable to the rollback vulnerability.
Note that we will remove all the \texttt{rem} instructions generated by EOSIO official libraries, e.g., eoslib, to reduce false positives.

\vspace{-0.1in}
\subsubsection{Missing Permission Check Detection}

As discussed in \S\ref{sec:vul:missing}, we focus on those functions that are valuable and lacking of authority validation before the sensitive operations.
Again, such functions should be reachable through the \texttt{apply} function. 
After filtering all the valuable functions according to criteria from \ref{eq:valuable:1} to \ref{eq:valuable:3}, we would examine if the constraints comply with the following conditions:

\vspace{-0.2in}
\begin{align*}
  (\{\forall func \in set_{func}, action \gets func\} \ &\cap \\
  \{code \gets \texttt{self}\}) \subseteq con_{apply} \ &\cap \\
  (\{\texttt{require\_auth}(acct)\} \centernot\subseteq con_{func})&
\end{align*} 

Specifically, the first condition firstly requires that the function (identified by $func$) is reachable by an attacker through the \texttt{apply} function.
Then, the second condition implies that the $func$ is allowed to be triggered directly. 
Lastly, the third condition checks if the $func$ lacks of authority verification.
If the above conditions are met, the smart contract is vulnerable to the missing permission check vulnerability.

%% file: Section-Implementation-Evaluation.tex
\section{Implementation and Experimental Setup}
\label{sec:evaluation}

\noindent \textbf{Implementation}
We take advantage of Octopus to construct the CFG of Wasm bytecode, and use the Z3 Theorem Prover (version 4.8.6) as our constraint solver. All the other major components, including Wasm Symbolic Execution Engine, Library Emulator and Vulnerability Scanners are all designed and implemented by ourselves. The implementation is based in Python, which includes over 5.5k lines of code.

\noindent \textbf{Experimental Setup}
Our experiment is performed on a server running Debian with four Intel(R) Xeon(R) E5-2620 v4 @ 2.10GHz and 64G RAM.
As mentioned in \S\ref{sec:tool}, the Wasm engine has
provided two configuration options (i.e., \textit{call depth}, and \textit{timeout}) to partially address the path explosion issue.
During our experiments, we empirically set the call depth as \textit{2 layers}, as we find it is enough to identify most vulnerabilities.
As to the exploration time, we empirically set the upper bound as \textit{5 minutes}, due to the following two main reasons.
First, within 5 minutes, all the smart contracts in our benchmark can be fully analyzed and detected with promising results (see \S\ref{sec:evaluation:rq1:benchmark}).
Second, as we seek to apply {\tool} to all the EOSIO smart contracts, we have to make a trade-off between accuracy and scalability. 
Therefore, the exploration time for each contract is set at a maximum of 5 minutes.
Note that, all these settings could be easily configured and customized in our tool, to fulfill the different requirements.

\noindent \textbf{Research Questions. }
Our evaluation is driven by the following three research questions (RQs).
\begin{itemize}
	\vspace{-0.1in}
	\item[RQ1] \textit{How accurate is {\tool} in detecting vulnerabilities of EOSIO smart contracts?} 
	\vspace{-0.1in}
	\item[RQ2] \textit{Are these vulnerabilities prevalent in the ecosystem?}
	\vspace{-0.1in}
	\item[RQ3] \textit{How many smart contracts have been exploited by attackers and what are the impacts of these attacks?} 
	\vspace{-0.1in}
\end{itemize}

To answer RQ1, in the absence of established benchmarks in the research community, we propose to collect real-world attacks and manually examine the victim smart contracts to craft a reliable benchmark. 
To answer RQ2, we collect all the available smart contracts on EOSIO and their historical versions. 
Then we apply {\tool} to detect the presence of security vulnerabilities, and characterize the evolution of vulnerabilities.
To answer RQ3, we further collect all the on-chain transactions related to the flagged vulnerable contracts, and then propose heuristics to pinpoint possible attacks.

\section{Experimental Results}

\subsection{RQ1: Accuracy of Vulnerability Detection}

\textbf{Benchmark.}
\label{sec:evaluation:rq1:benchmark}
To evaluate \tool, we first make efforts to craft a benchmark, which will be made available to the community.
EOSIO attacks were reported from time to time. 
Thus, we resort to the security reports released by well-known blockchain security companies to collect all the related publicly verified attacks~\cite{slowmist-zone, peckshield-blog} as the ground-truth. 
We have collected 38 attacks, targeting 34 unique vulnerable smart contracts in total.
Although these attacks were confirmed by the official team of the corresponding DApps, we found that some attacks are irrelevant to smart contract itself but other external factors, e.g., the server's issues~\cite{newdexpocket-attack}.
Thus, we further manually examined all the involved smart contracts.
Specifically, we found that 3 out of the 10 fake EOS attacks are related to server issues (e.g., \cite{newdexpocket-attack}). For rollback, 11 out of 21 attacks are due to the wrong reveal strategy of the server (e.g., \cite{betdiceadmin-rollback}). Besides, 2 of them were variants of rollback, which are related to the configuration of some nodes on EOS MainNet (see~\cite{rollback-blacklist}).
At last, we excluded all the above contracts from our benchmark to make sure all the attacks are resulted from the code in the smart contract itself.

The distribution of the benchmark is shown in Table~\ref{table:scanner-distribution-measurement}. 
Note that we also collected the corresponding patched smart contracts (without vulnerabilities) as comparison to evaluate the effectiveness of \tool. 
Additionally, there are only two vulnerable smart contracts related to the missing permission check vulnerability as reported and neither of them has been patched yet. 
Thus, we further manually created 4 pairs of smart contracts (with and without missing permission check vulnerability) to complement our benchmark. 
At last, we have labelled 52 smart contracts as our benchmark in total.

\begin{table}[!t]
\caption{Evaluation on the benchmark.}
\centering
\resizebox{\linewidth}{!}{

\begin{tabular}{r|c|ccc}
\hline
Vulnerability & \# Samples(Vul/Safe) & Precision & Recall  & F1-measure \\ \hline\hline
Fake EOS    & 14 (7/7)              & 100.00\%     & 100.00\% & 100.00\%    \\ \hline
Fake Receipt                                                       & 10 (5/5)              & 100.00\%     & 100.00\%   & 100.00\%      \\ \hline
Rollback                                                           & 18 (9/9)              & 100.00\%     & 88.89\% & 94.12\%    \\ \hline
\begin{tabular}[c]{@{}r@{}}Permission\end{tabular} & 10 (6/4)*             & 100.00\%     & 100.00\%   & 100.00\%      \\ \hline\hline
\textbf{Total}                                                      & 52 (27/25)              & 100.00\%     & 96.97\% & 98.46\%    \\ \hline
\multicolumn{5}{l}{* 4 pairs of the missing permission check samples are manually crafted.}                                \\
\end{tabular}
}
\vspace{-0.2in}
\label{table:scanner-distribution-measurement}
\end{table}

\textbf{Results.}
Among the 52 smart contracts, {\tool} flags 26 as vulnerable, with only one false negative case (belongs to rollback) and no false positives, leading to \textit{precision and recall of 100\% and 96.97\%}, respectively. 
Table~\ref{table:scanner-distribution-measurement} shows the detailed results.
For the only false negative case of rollback, i.e., \texttt{fairdogegame/betdogewallt}, the root cause is the number of suspicious \texttt{reveal} is too many to build path and symbolically execute each of them for a given \texttt{timeout} (\textit{5 minutes} here). After manually locating the vulnerable function, i.e., \texttt{func73}, we can get a correct result.
Therefore, the false negative is introduced by the optimization strategies, which is a trade-off between accuracy and scalability. 
It is easy to tune our approach to cover it, e.g., by exploring more paths and increasing the analyzing time.
Nevertheless, the exceptional case is rarely seen during experiments, as most smart contracts are not too complicated to handle.

\subsection{RQ2: Prevalence of Vulnerabilities}

\textbf{Dataset.}
We consider all the 53,666 smart contracts (including history versions) from June 9, 2018 (the very beginning of EOS MainNet) to November 15, 2019.
Note that, different from Ethereum smart contracts that cannot be modified once deployed, EOSIO contracts could be updated and bind with the same account as explained in \S\ref{sec:background:account}.
Thus, we use the \textit{EOSIO account} to label each \textit{unique smart contract}, i.e., one account may correspond to multiple \textit{contract versions}.
As a result, we have 53,666 different versions of contracts, which belong to 5,574 EOSIO accounts.
As the rollback vulnerability is only related to the gambling DApps, we can shrink our candidate list here. We refer to DAppTotal~\cite{dapptotal} -- a credible multi-platform DApp browser, to label the gambling DApps and use such contracts (17,394) for rollback vulnerability detection. 
For the other three kinds of vulnerabilities, we apply our detectors to all the 53,666 contracts (see Table~\ref{table:overall-results}).

\begin{table}[!t]
\caption{Vulnerability detection results in the wild.}
\centering
\resizebox{\linewidth}{!}{

\begin{tabular}{r|cc|cc}
\hline
Type                                                                 & \begin{tabular}[c]{@{}c@{}}\# Contracts\end{tabular} & \begin{tabular}[c]{@{}c@{}}\# Vulnerable (\%)\end{tabular} & \begin{tabular}[c]{@{}c@{}}\# Unique\end{tabular} & \begin{tabular}[c]{@{}c@{}}\# Vulnerable (\%)\end{tabular} \\ \hline\hline
Fake EOS & 53,666 & 1,457 (2.71\%) & 5,574 & 272 (4.88\%)  \\ \hline
Fake Receipt & 53,666 & 7,143 (13.31\%) & 5,574  & 2,192 (39.33\%) \\ 
\hline
Rollback & 17,394  & 1,149 (6.61\%)  & 913  & 84 (9.20\%)       \\ \hline
\begin{tabular}[c]{@{}r@{}}Permission\end{tabular} & 53,666  & 8,373 (15.60\%) & 5,574  & 662 (11.88\%) \\ \hline\hline
\textbf{Total} & \textbf{53,666} & \textbf{13,752 (25.63\%)}     & \textbf{5,574}  & \textbf{2,759 (49.50\%)} \\ \hline
\end{tabular}
}
\vspace{-0.2in}
\label{table:overall-results}
\end{table}

\vspace{-0.1in}
\subsubsection{Overall Results} 
Table~\ref{table:overall-results} shows the overall results. 
Surprisingly, over 25\% of the 53,666 smart contracts are vulnerable (see Column 3). The missing permission check vulnerability is the most prevalent, affecting over 15\% of the smart contracts. The fake receipt vulnerability is also quite common (13\%). 
For the rollback vulnerability, although we only analyzed 17K smart contracts of gambling DApps, over 1,000 of them are vulnerable.
The fake EOS vulnerability affects roughly 2.7\% of the smart contracts. \textit{It suggests that security vulnerabilities are prevalent in the EOSIO smart contract ecosystem, revealing the urgency to identify and prevent such vulnerabilities.}

\textbf{Vulnerable Unique Smart Contracts.}
As one smart contract may correspond to multiple versions, we further characterize the distribution of vulnerabilities from the perspective of unique contracts (accounts).
As shown in Column 5 of Table~\ref{table:overall-results}, for the 5,574 unique contracts, roughly half of them have at least one vulnerable version.
10\% of unique smart contracts account for 61.24\% of vulnerable versions, which indicates \textit{most of vulnerable versions are imported by a small portion of smart contracts.}
Besides, there are 1,793 unique smart contracts, whose versions are all vulnerable (41\% of them have at least two versions).
The contract \texttt{eossanguoone}, which is a popular game DApp, has the most number of vulnerable versions (356 versions).
By manual inspection, we found that all its versions released before September 4, 2019 have suffered from the fake receipt vulnerability, and then the vulnerability was patched by the developer.
The missing permission check vulnerability has been found since August 2019, which may be due to the importing of the new functions without authority check.

\begin{table}[!t]
\caption{The time to fix the vulnerabilities.}
\centering
\resizebox{\linewidth}{!}{

\begin{tabular}{r|cc|cc}
\hline
Type                                                                 & \# Unique (Vul) & \# Latest with Vul (\%) &\# Patched (\%) & Patch Time \\ \hline\hline
Fake EOS                                                             & 272                                                                                   & 207 (76.10\%)                                                                          & 65 (23.90\%)          & 14.85d                 \\ \hline
Fake Receipt                                                         & 2,192                                                                                 & 1,735 (79.15\%)                                                                        & 457 (20.85\%)          & 24.01d                 \\ \hline
Rollback                                                             & 84                                                                                    & 28 (33.33\%)                                                                           & 56 (66.67\%)          & 4.24d                  \\ \hline
\begin{tabular}[c]{@{}r@{}}Permission\end{tabular} & 662                                                                                   & 313 (47.28\%)                                                                          & 349 (53.72\%)          & 4.38d*                  \\ \hline\hline
\textbf{Total}                                                       & \textbf{2,759}                                                                        & \textbf{2,080 (75.39\%)}                               & \textbf{679} (\textbf{24.61\%}) & \textbf{16.84d}        \\ \hline
\multicolumn{5}{l}{\begin{tabular}[c]{@{}l@{}}*The average patch time for missing permission check is calculated on the action level.\end{tabular}}               \\
\end{tabular}
}
\vspace{-0.2in}
\label{table:latest-vul}
\end{table}

\vspace{-0.1in}
\subsubsection{Time to fix the vulnerability}
\label{subsubsec:time-fix}
As we have analyzed the evolution of vulnerabilities across different versions, it is thus necessary to further investigate the \textit{time to fix the vulnerabilities} for each unique smart contract, which could be used to measure the \textit{window period} for the attackers to exploit these vulnerabilities.

\textbf{Result.} 
As shown in Table~\ref{table:latest-vul}, for the 2,759 unique smart contracts with vulnerable versions, over 75\% of them still have at least one security vulnerability in their latest version by the time of our study. 679 unique smart contracts have patched all their vulnerabilities during their evolution, and the average window period is 16.84 days.

\textbf{Patch Rate.}
We further analyze the patch rate across vulnerabilities.
The rollback vulnerability has the highest patch rate (over 66\%), and the average window period is roughly 4 days. 
The reason for its timely response might be that the rollback vulnerability only exists in game/gambling DApps, which usually have high balance in their accounts. The financial loss could be devastating if developers leave the vulnerability alone.
For the missing permission check, 
349 smart contracts have patched all their missing check actions. 
Note that, we measured the average patch time on the action level here, as one vulnerable contract may have more than one missing permission check actions. There are 647 patched actions in total -- roughly 500 of them are patched within only one day, while the overall patch time is 4.38 days.
It suggests that most of the missing permission checking actions are patched timely, while a few contracts take relative long time to fix.
In contrast, the fake EOS and the fake receipt vulnerabilities have the lowest patch rates (i.e., roughly 20\%), and the patching time is relative long (i.e., 2 to 3 weeks on average). Our manual check found that, half of the smart contracts related to fake receipt are patched within 24 hours, which further indicates that some \textit{inactive} smart contracts drag the average patch time.
Most of the inactive smart contracts (accounts) have no balance and very few transactions, which are usually not the targets of attackers.

\vspace{-0.1in}
\subsection{RQ3: The Presence of Attacks}
\label{subsec:attack}

\subsubsection{Approach}
It is not trivial to explore how many of the vulnerable smart contracts have been successfully exploited by the attackers. Until recently, a lot of ad hoc (often manual) efforts of security researchers~\cite{peckshield-blog, slowmist-zone} are necessary to verify them.
Thus, given the vulnerable smart contracts, we first collect all their related \textit{on-chain} transactions, and then design a set of heuristics to locate the suspicious attacks, which will be used to facilitate further manual verification to determine the real attacks. In total, we have collected over $2.5 billion$ transaction records.

\textbf{Fake EOS Attack.} 
The most important behavior of this attack is to defraud the \textit{official EOS tokens} from the vulnerable smart contract by using the \textit{fake EOS tokens}, which can be identified through the transaction records storing the information of token issuers. 
According to the observation, we will first filter out all the transactions of token transfer whose token symbols are ``EOS''.
Then, these transactions will be grouped according to the following definitions:
\begin{itemize}
    \vspace{-0.1in}
    \item \textit{fake-sending} transactions that send fake EOS tokens.
    \vspace{-0.1in}
    \item \textit{true-sending} transactions that send true EOS tokens.
    \vspace{-0.1in}
    \item \textit{true-receiving} transactions that receive true EOS tokens.
    \vspace{-0.2in}
\end{itemize}
As a result, we can define a \textit{potential} attack as a sequence of a fake-sending transaction followed by a true-receiving transaction. Note that a fake-sending transaction A can be joined with a true-receiving transaction B, if and only if they appear on the same period while A occurs before B. 
For all these potential transactions, we focus mainly on those who have gained more true EOS tokens than they spent. To this end, we further examine the input-output ratio between the attacker and the vulnerable contracts to determine the \textit{suspicious} attacks.
Finally, based on the suspicious attacks, we will verify whether the vulnerable smart contracts will resume the normal execution (e.g., running a lottery for a real player) after receiving the fake EOS tokens. If so, we will mark the suspicious transaction as a fake EOS attack.

\textbf{Fake Receipt Attack.} 
The key feature of this attack is that the vulnerable smart contract is misled by the fake notification to receive tokens, while the actual token transfer occurs between the two accounts belonging to the same attacker (see \S\ref{sec:vul:fake-receipt}). For simplicity, we will use \textit{from\_account} and \textit{to\_account} to represent the two accounts in the following, where {to\_account} will send the fake receipt to vulnerable contract, and {from\_account} is the ultimate beneficiary.  

Accordingly, we will first query all the transactions of token transfer whose tokens are issued by \texttt{eosio.token} and token symbols are ``EOS'', to get all the true EOS token transfers.
Then, we will filter out the transactions whose receivers are neither \texttt{eosio.token}, nor the {from\_account} or {to\_account}. These transactions will be regarded as the fake receipts with crafted notifications.
Next, if a {from\_account} sends a fake receipt before making profits from the vulnerable contract, we will mark the corresponding transaction as potential. 
After that, by eliminating the unrelated EOS spending transactions (e.g., for testing purpose initiated by the attacker), we focus mainly on those who have gained more true EOS tokens than they spent. If the input-output ratio are still high, the corresponding transactions are labeled as \textit{suspicious}.

Finally, we will manually check the suspicious transactions whether the vulnerable smart contract will resume the normal execution after receiving the fake receipts. If so, we will mark such a transaction as a fake receipt attack.
    
\textbf{Rollback Attack.} 
As mentioned in \S\ref{sec:vul:rollback}, the transaction of this attack is composed of sequential invocations of actions, which can be used as the pattern to identify the attack.

Specifically, we will first filter out all the transactions who contains at least four actions as the potential transactions. 
Next, we will select suspicious transactions which as long as meet the following four conditions: 
(1) the first and the last actions must be invoked in the same contract, where the first means to start the attack, and the last will determine whether the rollback is necessary after receiving the reward from the vulnerable smart contract.
(2) the two actions in the middle must be token transfers through \texttt{eosio.token}, and the sender and the receiver (either one must be the vulnerable smart contract) of the two actions are arranged opposite to each other.
(3) at least one of the counterparties, i.e., either the sender or the receiver, is labeled as the gambling or game DApp.
(4) the amount of tokens transferred from the vulnerable smart contract is more than it received.
Besides, it is worth noting that, the rollbacked transactions will not be recorded on the chain. 
As a result, we have to manually check the player's successful rate per unit time, namely, if it is oddly high than the others, we will mark the suspicious transaction as a rollback attack.

\textbf{Missing Permission Check Attack.} 
Because authority information is along with the invoked transaction, we can examine whether it belongs to the callee contract to identify this attack.
More precisely, we will first screen out all the transactions whose target actions are the vulnerable actions, to get suspicious transactions.
Then, if the transaction's authority does not belong to that smart contract the action belongs to, we will mark it as a missing permission check attack.
\vspace{-0.1in}
\subsubsection{Results}

\begin{table}[!t]
\caption{Overall results of attack detection.}
\centering
\resizebox{\linewidth}{!}{
\begin{tabular}{r|ccc|c}
\hline
Type                                                                 & \# Attacks   & \begin{tabular}[c]{@{}c@{}}\# Attacker\\ / Victims\end{tabular} & Financial Loss (\$)   & \# Verified \\ \hline\hline
Fake EOS                                                             & 9                    & 10 / 9               & 652,428.48             & 8           \\ \hline
Fake Receipt                                                         & 27                   & 28 / 17              & 1,020,831.94           & 7           \\ \hline
Rollback                                                             & 12                   & 12 / 9               & 52,984.00              & 12          \\ \hline
\begin{tabular}[c]{@{}r@{}}Permission\end{tabular} & 183                  & - / 144              & -                      & -           \\ \hline\hline
\textbf{Total}                                                       & \textbf{48*}         & \textbf{50 / 34*}    & \textbf{1,726,244.42}  & \textbf{27} \\ \hline
\multicolumn{5}{l}{* Exclude the results of missing permission check.}                                \\
\end{tabular}
}
\vspace{-0.2in}
\label{table:overall-results-history-attacks}
\end{table}

\begin{table}[!t]
\caption{Top 5 identified attack events.}
\centering
\resizebox{\linewidth}{!}{

\begin{tabular}{r|cc|cc}
\hline
Attack Type                   & Attacker Account(s)       & Victim Account                & EOS Loss            & Amount Loss (\$)           \\ \hline\hline
Fake Receipt                  & il***23, wh***r1          & eosbetdice11                  & 138,319.80          & 756,609.30            \\ \hline
Fake EOS                      & re***et                   & eoscastdmgb1                  & 63,014.10           & 327,673.32            \\ \hline
Fake Receipt                  & re***om, re***et          & nkpaymentcap                  & 53,641.71           & 200,619.98            \\ \hline
Fake EOS                      & aa***fg                   & eosbetdice11                  & 44,427.43           & 234,132.56            \\ \hline
Fake Receipt                  & be***s1, be***s2          & epsdcclassic                  & 17,388.73           & 41,559.07             \\ \hline\hline
\textbf{Total}                    & \textbf{-}                & \textbf{-}                    & \textbf{341,437.30} & \textbf{1,638,803.89} \\ \hline
\end{tabular}
}
\vspace{-0.2in}
\label{table:top10-history-attack}
\end{table}

The overall result is shown in Table~\ref{table:overall-results-history-attacks}. We have identified 48 attacks in total, including 9 fake EOS attacks, 27 fake receipt attacks, and 12 rollback attacks. Additionally, we also identified 183 invoked actions (belonging to 144 contracts) which missed the permission check (see Table~\ref{table:missing-permission-check-history} in Appendix).
Note that, for these missing permission check actions, some of them are designed intentionally instead of unexpected implementation. It is hard to differentiate whether they are attacks or not, and it is impossible to estimate the financial loss. Therefore, we regard them as misuse actions instead of attacks.

\textbf{Impact of Attacks.}
\textit{The 48 identified attacks lead to over 341K EOS loss, which is roughly 1.7M USD according the close price of the date of attacks.}    
Note that we have collaborated with a leading blockchain security company to report these attacks to the DApp developers, and 27 of them have been confirmed, accounting for more than 99\% of the total loss. All the unconfirmed suspicious attack events only relate to a few EOS, and most of them are no longer active.
The Top-5 confirmed attack events are listed in the Table~\ref{table:top10-history-attack}.

\textbf{Unexploited Vulnerable Contracts.}
It is interesting to observe that, although thousands of contracts are vulnerable (see Table~\ref{table:overall-results}), only a few of them have been successfully exploited by attackers.
We manually sampled some smart contracts for reverse engineering and inspecting their transactions, we found there are mainly two reasons leading to this.
First, the popular smart contracts (with high balance) are the main targets of attackers, but these vulnerable contracts can be patched in time, and leave a very short window period for attackers. 
Based on the transactions, we found that attackers are always trying to exploit the popular contracts, and some attacks are indeed succeed (see Table~\ref{table:top10-history-attack}), while most of them were failed.
Second, as we mentioned in \S\ref{subsubsec:time-fix}, most of the unpatched smart contracts are inactive ones that have low balance, which attract low attention of attackers, considering the trade-off between low profit and the cost of attacks.

%% file: Section-Discussion.tex
\vspace{-0.1in}
\section{Threats to Validity}
\label{sec:discussion}

First, \emph{our system inherits the limitation of symbolic execution}, i.e., path explosion. Although we have implemented several optimization strategies, {\tool} still reports false negative case, as discussed in \S\ref{sec:evaluation:rq1:benchmark}.
However, this is not a big issue for our system.
On one hand, most of the smart contracts are not as complicate as other software. A large portion of smart contracts can be fully analyzed in a short time. On the other hand, we have proposed specific optimization methods when searching for the vulnerabilities that could eliminate most irrelevant paths.
Nevertheless, we agree that we can take advantage of advanced symbolic execution techniques~\cite{godefroid2008automated, burnim2008heuristics, cadar2008klee, collingbourne2014symbolic, sen2015multise, trabish2018chopped} to alleviate this issue.
Second, \emph{we rely on heuristics and semi-automated methods to verify attacks} (see \S\ref{subsec:attack}). This, of course, might not be scalable and could mean that we only offer a lower-bound of attacks.
However, a large portion of the attacks we identified are confirmed by DApp teams, which suggests that our approach is quite reliable. Nevertheless, we agree that some techniques (e.g., dynamic testing) can be applied to help us automatically identify attacks. 
In this paper, our main contribution is automatically detecting the security vulnerabilities, while attack verification is not a main focus in this work.
Third,
\emph{there might be some new vulnerabilities we did not cover in this current prototype, as well as the general vulnerabilities in other software systems}, such as buffer overflow.
In this paper, we focus only on the EOSIO-specific vulnerabilities, the main reason is that we are lacking ground-truth of other security bugs. 
Nevertheless, we argue that it is easy to extend our system to cover other vulnerabilities, as the symbolic execution engine and the scanner framework are generic. Moreover, {\tool} can also work on Wasm bytecode of other platforms (e.g., web), while the only effort is to resolve the library dependency of the corresponding platform.

%% file: Section-Related.tex
\vspace{-0.1in}
\section{Related Work}

\textbf{WebAssembly Bytecode Analysis}
WebAssembly is the new low-level language for the web.
There are only a handful work on analyzing the Wasm bytecode~\cite{szanto2018taint, fu2018taintassembly, lehmann2019wasabi, disselkoen2019position, vassena2019memory}. 
For example, Lehmann et al.~\cite{lehmann2019wasabi} has proposed a general-purpose dynamic analysis system for Wasm, which allows developers or researchers to implement heavyweight dynamic analysis, e.g., instruction counting and memory access tracing. 
However, all of them were focused on web applications, which were mainly dynamic analysis.
In this paper, we implemented a general symbolic execution framework for Wasm, and made effort to support the security analysis of EOSIO smart contracts.

\textbf{EOSIO Analysis}
There are several works were focused on the EOSIO ecosystem~\cite{bach2018comparative, xu2018eos, min2019security, huang2020characterizing, lee2019spent}. 
For example, Huang et al.~\cite{huang2020characterizing} proposed a method to identify the bot-like accounts in EOSIO based on transaction analysis.
Lee et al.~\cite{lee2019spent} introduced and studied four attacks stemming from the unique design of EOSIO.
Several technical blogs~\cite{eosbet-attack-fake-receipt, eoscast-attack, transaction-congestion, eosbet-attack-fake-eos, newdexpocket-attack, rollback-blacklist} from the industry have reported the security attacks of EOSIO.
However, there are no available work available on detecting the security vulnerabilities in EOSIO.

\textbf{Vulnerability Detection of Ethereum Smart Contracts}
Ethereum has received lots of attention from academia, and a number of studies were focused on the vulnerability detection~\cite{jiang2018contractfuzzer, tikhomirov2018smartcheck, grech2018madmax, krupp2018teether, torres2018osiris, hildenbrandt2018kevm, luu2016making, he2019characterizing, wang2020contractward, liao2019soliaudit, gao2019easyflow, lai2020static}.
For example, \cite{torres2018osiris, gao2019easyflow, lai2020static} were mainly focused on the overflow vulnerabilities. 
Luu et al.~\cite{luu2016making} proposed Oyente, the first symbolic execution tool for detecting vulnerabilities in Ethereum smart contracts.
Machine learning and fuzz testing techniques \cite{wang2020contractward, liao2019soliaudit} were also adopted to identify the vulnerabilities of Ethereum smart contracts.
As we mentioned earlier, \textit{the two ecosystems (Ethereum and EOSIO) are totally different, and no previous work on Ethereum can be applied to analyze EOSIO smart contracts directly.} 
Nevertheless, we admit that the general idea of Ethereum vulnerability detection can be used to improve our work.

%% file: Section-Appendix.tex
\section{Appendix}
\vspace{-0.5in}

$\,$

\subsection{Missing Permission Check Actions.}
\vspace{-0.1in}
We have identified 183 missing permission check actions, as shown in 
Table~\ref{table:missing-permission-check-history}.

\begin{table*}[htb]
\vspace{-0.5in}
\caption{A List of Missing Permission Check Actions and their Corresponding Accounts.}
\centering
 \resizebox{0.6\textwidth}{!}{

\begin{tabular}{|cc|cc|cc|}
\hline
Account      & Action       & Account      & Action       & Account      & Action       \\ \hline\hline
214odicedice & leave        & eosknightsio & rebirth3     & oneplayslots & clearrow     \\ \hline
akdexiononce & clear        & eosknightsio & skcsell      & oyeyeyeyeyeo & idex         \\ \hline
akdexiononce & nonce        & eosknightsio & signup       & oyeyeyeyeyeo & iban         \\ \hline
alibabapool1 & jackpot      & eosknightsio & skillreset   & paritysupply & adduser      \\ \hline
aqsensordata & update       & eosknightsio & petgacha3    & paritysupply & newaccount   \\ \hline
arbarotokenn & create       & eosknightsio & itemlvup3    & paritysupply & stake        \\ \hline
arbarotokenn & claim        & eosknightsio & rebirth      & parslseed123 & refund       \\ \hline
baccarat.e   & reveal       & eoslotteryes & del          & pcscoreprtcl & refreshkey2  \\ \hline
bairenniuniu & reveal       & eospayserver & login        & pcscoreprtcl & refreshkey   \\ \hline
bancor3dcode & jackpot      & eospredictio & reqpredict   & pickownbonus & withdrawown  \\ \hline
battlebricks & startattend  & eosramoption & trigger      & pickowngames & withdrawref  \\ \hline
betmoonadmin & makebet      & eossanguoone & operateboss  & pinganwallet & m            \\ \hline
bingobetgame & playlucky    & eoswinnerdic & initcontract & pntvuxfbguce & clear        \\ \hline
bingobetgame & playbonuslot & eosyxtoken22 & unstake      & pptqipae1yog & m            \\ \hline
blackjack.e  & resolve      & eosyxtoken22 & getshare     & pythongolang & update       \\ \hline
blockfishbgp & adrequest    & eparticlectr & brainclmid   & rabgamecoins & login        \\ \hline
bluebetthree & login        & eparticlectr & fnlbyhash    & rabgamecoins & logingame    \\ \hline
bluebetthree & firstlogin   & eparticlectr & oldvotepurge & rating.pra   & check        \\ \hline
bosibc.io    & rmunablerb   & eparticlectr & finalize     & resetcontrac & childreflect \\ \hline
bosibc.io    & rollback     & eparticlectr & rewardclmid  & romangame222 & loginvestor  \\ \hline
bosibc5chain & rmfirstsctn  & eparticlectr & procrewards  & roulette.e   & reveal       \\ \hline
candy.w      & manureward   & exchangename & accomplish   & sanlijishubu & sign         \\ \hline
casinolordio & withdraw     & fairkuai3kkk & close        & scgspzuufcce & rand         \\ \hline
conquerworld & end          & farmeosrich1 & endprofit    & scratchcards & scratch      \\ \hline
crheroestest & battle       & farmeosrich1 & profit       & scratchers55 & reveal       \\ \hline
cryptsangoku & clear        & farmeosrichx & profit       & signupeoscom & clearexpired \\ \hline
daccustodia1 & newperiod    & farmeosrichx & endprofit    & slotcontract & initstat     \\ \hline
dappbaccarat & reveal       & findexfindex & executetrade & slotmachine1 & reveal       \\ \hline
dappshieldio & addsdkconfig & g4ydgmjyhege & deleterow    & stakemine123 & refresh      \\ \hline
deltadexcode & preparetrade & g4zdkobqhege & deleterow    & string.x     & startgame    \\ \hline
dgatepokergc & reset        & gamblrprofit & unstake      & superarmy123 & upduser      \\ \hline
dgatepokergc & resetladder  & gameskypools & release      & testblueuser & updateprofit \\ \hline
dgatepokerpr & stake        & godice.e     & reveal       & thebetxowner & printresults \\ \hline
dicestaker5a & verifyuser   & gopokerdotio & verifycards  & thedeosgames & stake        \\ \hline
dollarbillgo & check        & gyftietokgen & generate     & thedepositgw & mint         \\ \hline
dsdaeveafaef & claim        & helloworldjs & out          & therealkarma & refund       \\ \hline
dslots123123 & resolve      & horustokenio & claimreward  & tothemoonmnt & cbldng       \\ \hline
dtheoschain1 & deleteroom   & horustokenio & refundbyid   & trustbetchat & reveal       \\ \hline
eegg.io      & resetcounter & horustokenio & refundhorus  & trybenetwork & addtester    \\ \hline
elpaymentcom & claim        & ilove.q      & go           & tttblackjack & dg           \\ \hline
enserve.bank & migdata      & jmihongbao11 & recv         & ultrahikkash & withdraw     \\ \hline
eocfexchange & ugo          & kdsrpgkdsrpg & fabi         & untdevtooth1 & deletesys    \\ \hline
eosbankgamea & prepare      & koi111111111 & bid          & untowertest1 & deletemail   \\ \hline
eosbanksfund & prepare      & kuai3iostake & unstake      & usercontract & buy          \\ \hline
eosbaoserver & jackpot      & kuai3iostake & refund       & virtualusers & checksign    \\ \hline
eoscubetoken & signup       & lumeospollss & uppolllikes  & wangshaoyong & setcfg       \\ \hline
eosdaqonswap & eraseconfig  & lumeospollss & upcmntlikes  & warofstar.e  & reveal       \\ \hline
eosdlongjohn & gen          & lumetokenctr & lock         & wealthplan33 & fuverify     \\ \hline
eosdtcntract & positionadd  & lumetokenctr & unlock       & weosservices & destroytoken \\ \hline
eosdtorclize & refreshutil  & lynxeosgame2 & cleardb      & whaleexgate4 & execute      \\ \hline
eosevenstake & staking      & lynxeosgame3 & setpubkey    & whaleexhelpu & unstake      \\ \hline
eosevenstake & claim        & lynxeosgame3 & cleardb      & whaleexhelpu & stake        \\ \hline
eosfakerbatl & init         & lynxeosgame4 & cleardb      & wizardfights & cleanwizcd   \\ \hline
eosgamesprod & pong         & marvellous3d & reveal       & wizardmarket & createsale   \\ \hline
eosioshadows & jackpot      & mhmttestcont & addteainfo   & wizardstoken & createwizard \\ \hline
eosjackypool & claim        & monstereosio & feedpet      & worldconques & end          \\ \hline
eoskeydice11 & initcontract & mymillionsio & collect.one  & xiongzhend13 & replacebet   \\ \hline
eosknightsio & skwear       & mymillionsio & collect.all  & xmassnowball & updatecamp   \\ \hline
eosknightsio & sksell       & nkpaybankcap & updateclaim  & xpetiocore11 & ping         \\ \hline
eosknightsio & itemlvup     & nkpaybankcap & claim        & yizeshenzhen & eraseconfig  \\ \hline
eosknightsio & petgacha     & oneplaygames & clearrow     & yumsactivity & hit          \\ \hline
\end{tabular}
 }
\label{table:missing-permission-check-history}
\end{table*}